\documentclass[a4paper]{jpconf}
\usepackage{graphicx}
\usepackage{amsmath,amssymb,amsfonts}

\usepackage{setspace}
\usepackage[T1]{fontenc}
\usepackage[latin1]{inputenc}
\usepackage{epsfig}
\usepackage[english]{babel}
\usepackage{color}
\usepackage{dcolumn}
\usepackage{moreverb}
\usepackage{times}
\usepackage{palatino} 
\usepackage{lmodern}
\usepackage[all]{xy}

\begin{document}

\title{New families of superintegrable systems from k-step rational extensions, polynomial algebras and degeneracies}

\author{Ian Marquette}

\address{School of Mathematics and Physics, The University of Queensland, Brisbane, QLD 4072, Australia}

\ead{i.marquette@uq.edu.au}

\begin{abstract}
Four new families of two-dimensional quantum superintegrable systems are constructed from k-step extension of the harmonic oscillator and the radial oscillator.  Their wavefunctions are related with Hermite and Laguerre exceptional orthogonal polynomials (EOP) of type III. We show that ladder operators obtained from alternative construction based on combinations of supercharges in the Krein-Adler and Darboux Crum ( or state deleting and creating ) approaches can be used to generate a set of integrals of motion and a corresponding polynomial algebra that provides an algebraic derivation of the full spectrum and total number of degeneracies. Such derivation is based on finite dimensional unitary representations (unirreps) and doesn't work for integrals build from standard ladder operators in supersymmetric quantum mechanics (SUSYQM) as they contain singlets isolated from excited states. In this paper, we also rely on a novel approach to obtain the finite dimensional unirreps based on the action of the integrals of motion on the wavefunctions given in terms of these EOP. We compare the results with those obtained from the Daskaloyannis approach and the realizations in terms of deformed oscillator algebras for one of the new families in the case of 1-step extension. This communication is a review of recent works.
\end{abstract}

\section{Introduction}

Superintegrable systems in classical and quantum mechanics have many properties that make them very interesting models from both mathematics and physics point of view \cite{miller13}. Among these properties, the algebra of integrals of motion is non-Abelian and usually it is a finitely generated polynomial algebra. Moreover, this polynomial algebra in quantum mechanics provides energy spectra and information on wave functions and leads to an additional degeneracy of energy levels. This is however a quite complicated task for systems with higher order integrals to obtain the polynomial algebra generated by the integrals even in two-dimensions and in the case of cubic algebras \cite{marquette09a}. Moreover, the analysis of the representations in order to obtain an algebraic derivation of the energy spectrum and total number of degeneracies is also a difficult problem. A method was introduced by Dakaloyannis \cite{daska01} to study representations of quadratic algebras based on Casimir operators and realizations as deformed oscillator algebras and was applied on Smorodinsky-Winternitz systems \cite{win67}. This approach was generalized to cubic algebras \cite{marquette09a} and even to a class of polynomial algebras of arbitrary order with three generators \cite{Isaac14}. 
\newline
\newline
It has been pointed out that not all the spectrum and the degeneracies were recovered from the finite dimensional unitary representations (unirreps) \cite{marquette09a} for many of the Gravel systems \cite{gra04}. More recently, we showed that these systems are first members of families of superintegrable systems involving 1-step extension of harmonic oscillator \cite{marquette13a} and related to Hermite exceptionnal orthogonal polynomials (EOP) of type III ($X_{m}$). Such orthogonal polynomials were introduced quite recently in context of Sturm-Liouville problems and supersymmetric quantum mechanics (SUSYQM) \cite{gomez10a,gomez09,gomez13b,gomez13b,cq08,cq11,oda10,oda11} but the connection with superintegrability was much more unexplored and only one example related to Jacobi EOP was studied \cite{post12}. It was pointed out that the incomplete algebraic solutions were due to the construction of integrals from standard ladder operators in supersymmetric quantum mechanics and that these ladder operators admit isolated singlets from excited states \cite{marquette13a}. However, all the energy levels and degeneracies were obtained algebraically for the 1-step extension of radial oscillator related Laguerre EOP of type I and II, but not for the type III. It was demonstrated in another paper that new type of ladder operators with $m+1$ infinite chains of levels can be created in the case of 1-step extension of the harmonic oscillator from which a new set of integrals and a corresponding polynomial algebra provide via finite dimensional unirreps the whole spectrum and total number of degeneracies \cite{marquette13b}. We also pointed out that another type of ladder operators with doublet states and many infinite chains of levels \cite{marquette13c} for 2-step extension of harmonic oscillator exists, which showed that ladder operators for a given systems are far than unique. However, these new ladder operators for these 1D exactly solvable systems related to type III Hermite EOP ($X_{m_{1},m_{2}}$) do not allow to generate integrals from which all the spectrum can be obtained for 2D superintegrable generalisations. From these results, we presented a way to construct new ladder from deleting and creating approach of supersymmetric quantum mechanics for k-step extension of the harmonic oscillator and radial oscillator both related to type III EOP ( Hermite and Laguerre ). We constructed new superintegrable models and showed how from a new set of integrals and its corresponding polynomial algebra a complete algebraic derivation of the whole spectrum and total number of degeneracies can be performed using a novel approach based on the action of the integrals \cite{marquette14a}. 
\newline
The purpose of this communication is to present a review of some of these results obtained in collaboration with C.Quesne \cite{marquette13a,marquette13b,marquette13c,marquette14a} and point out also the more complex patterns of the energy levels and unirreps for these superintegrable Hamiltonians.
\newline
The paper is organized as follow. In Section 2, we present four families of 2D superintegrable systems and their degeneracies. In Section 3, we present the polynomial algebras of these systems. In Section 4, we present an algebraic derivation of the degeneracies of the complete spectrum using a direct approach for the four cases involving k-step extension. In Section 5, we compare these results obtained in Section 4 in the case of 1-step for one of the families using Daskaloyannis approach to obtain finite dimensional unirreps.

\section{ New superintegrable systems and degeneracies}

In previous paper we introduce new families of quantum superintegrable systems from k-step extension of the harmonic oscillator
\begin{equation}
\begin{split}
 H_{a} &= H_{x}^{(ko)}+H_{y}^{(o)}= - \frac{d^2}{dx^2} - \frac{d^2}{dy^2} + x^2  + y^2 - 2k  \\
& \quad - 2 \frac{d^2}{dx^2}   \log {\cal W}({\cal H}_{m_1}, {\cal H}_{m_2}, \ldots, {\cal H}_{m_k}), \label{Ha}
	\end{split}
\end{equation}
\begin{equation}
\begin{split}
  H_{b} &= H_{x}^{(kro)}+H_{y}^{(o)}=  - \frac{d^2}{dx^2} - \frac{d^2}{dy^2} + \frac{1}{4} x^2 + \frac{l(l+1)}{x^2} + y^2 - k \\
  & \quad - 2 \frac{d^2}{dx^2} \log \tilde{\cal W}\bigl(L^{(-\alpha-k)}_{m_1}(-z), L^{(-\alpha-k)}_{m_2}(-z), \ldots,
       L^{(-\alpha-k)}_{m_k}(-z)\bigr) , \label{Hb}
\end{split}
\end{equation}
\begin{equation}
\begin{split}
  H_{c}  &= H_{x}^{(ko)}+H_{y}^{(ro)}= - \frac{d^2}{dx^2} - \frac{d^2}{dy^2}+ x^2 + \frac{1}{4} y^2 + \frac{l(l+1)}{y^2} - 2k \\
   & \quad  - 2 \frac{d^2}{dx^2} \log {\cal W}({\cal H}_{m_1}, {\cal H}_{m_2}, \ldots,{\cal H}_{m_k}), \label{Hc}
\end{split}
\end{equation}
\begin{equation}
\begin{split}
  H_{d} &=  H_{x}^{(kro)}+H_{y}^{(ro)}=  - \frac{d^2}{dx^2} - \frac{d^2}{dy^2}+ \frac{1}{4} x^2 + \frac{l(l+1)}{x^2} +  \frac{1}{4} y^2 + \frac{l(l+1)}{y^2} - k \\
  & \quad - 2 \frac{d^2}{dx^2} \log \tilde{\cal W}\bigl(L^{(-\alpha-k)}_{m_1}(-z), L^{(-\alpha-k)}_{m_2}(-z), \ldots, L^{(-\alpha-k)}_{m_k}(-z)\bigr) , \label{Hd}
\end{split}
\end{equation}
where the 1D components $H^{(o)}$, $H^{(ro)}$, $H^{(ko)}$ and $H^{(kro)}$ in Eqs.~(\ref{Ha}),~(\ref{Hb}),~(\ref{Hc}) and~(\ref{Hd}) are respectively the harmonic oscillator, the radial oscillator, the k-step extension of the harmonic oscillator and the k-step extension of the radial oscillator \cite{marquette14a}. We have the condition $\alpha + k > m_k + 1$ and ${\cal H}_m(x) = (-{\rm i})^m H_m({\rm i}x)$ is a $m$th-degree pseudo-Hermite polynomial and $L^{(\alpha)}_{\nu}(z)$ denotes a $\nu$th-degree Laguerre polynomial. $m_1 < m_2 < \cdots < m_k$ with $m_i$ even (resp.\ odd) for $i$ odd (resp.\ even).

For convenience, the parameter $\hbar$ have been set to 1 in all the quantum Hamiltonians, supercharges, ladder and integrals \cite{marquette14a}, however the classical limit $\hbar \rightarrow 0$ of systems $H^{(ko)}$ and $H^{(kro)}$ will be only the harmonic oscillator and the radial oscillator. This is a consequence that these k-step extensions and their ladder operators are build from supercharges in supersymmetric quantum mechanics which is by essence a quantum formalism. However, ladder operators and related deformed Heisenberg algebras can still be defined  in classical mechanics by replacing the commutator by the Poisson bracket and obtained by other methods \cite{kuru08,marquette12a} as SUSYQM can not be used to generate such operators. However, as for the case of superintegrable systems, the search and classification of 1D classical systems with analog of ladder operators needs to be performed separately than the quantum case, but such 1D systems can also be used to generate 2D classical superintegrable systems \cite{marquette12a}.

The wave functions can be calculated using Darboux-Crum or Krein-Adler formalism and are related to Hermite and Laguerre EOP of type III. These four families of systems generalise the Smorodinsky-Winternitz Hamiltonians \cite{win67}. Moreover, it can be shown using the separation of variables that the energy spectrum $E_{i,N}$ of the corresponding Hamiltonian $H_{i}$ ($i=a,b,c,d$) is
\begin{equation}
  E_{i,N}=2N + \gamma_{i},\quad N=\nu_{x}+\nu_{y}+1, \label{Ener}
\end{equation}
where the constants $\gamma_{a}=0$, $\gamma_{b}=\alpha+k$, $\gamma_{c}=\alpha$, $\gamma_{d}= 2 \alpha + k$, and the quantum numbers $\nu_{x}$ and $\nu_{y}$ can take the following values
\begin{equation}
 \nu_{x}=-m_{k}-1,\ldots,-m_{1}-1,0,1,2,\ldots, \qquad \nu_{y}=0,1,2\ldots. 
\end{equation}
A direct calculation of the degeneracies of the spectrum given by Eq.~(\ref{Ener}) by an enumeration process leads to the following pattern
\begin{equation}
  \deg(E_{N}) = \begin{cases}
      k-j+1  & \text{if $N=-m_{j},-m_j+1,\ldots,-m_{j-1}-1$}, \\ 
        & \text{for $j=2,3,\ldots,k$}, \\ 
      k  & \text{if $N=-m_{1},-m_1+1,\ldots,0$}, \\ 
      N+k  & \text{if $N=1,2,3,\ldots$}. \\  \label{deg}
  \end{cases} 
 \end{equation}
This is one of the distinctive feature of these superintegrable Hamiltonians to exhibit more complicate patterns for the degeneracies given by Eq.~(\ref{deg}) than what was previously observed in other models \cite{miller13}. 
Let us emphasis that in other superintegrable Hamiltonians obtained even if the finitely generated polynomial algebra contains higher order terms and even of arbitrary order as for the 2D Tremblay-Turbiner-Winternitz (TTW) systems \cite{trem09,miller13,kal11,kal12}, the degeneracies are similar to the 2D anisotropic harmonic oscillator. We thus need to provide an algebraic explanation for unusual degeneracies in Eq.(7) of these four quantum systems $H_{a}$, $H_{b}$, $H_{c}$ and $H_{d}$ i.e. we need to obtain the whole energy spectrum and total number of degeneracies from the unirreps of the polynomial algebra and the Section 3, 4 and 5 of this papers are devoted to this task.

\section{Construction of the integrals and the polynomial algebras}

These four families of systems presented in previous section and given by Eqs.~(\ref{Ha}),~(\ref{Hb}),~(\ref{Hc}) and~(\ref{Hd}) belong to the following class of Hamiltonian that consists in sum of two 1D Hamiltonians and allowing separation of variables in Cartesian coordinates 
\begin{equation}
  H=H_{x}+H_{y}=-\frac{d^{2}}{dx^{2}}-\frac{d^{2}}{dy^{2}}+V_{x}(x)+V_{y}(y), \label{hamil}
\end{equation}
and allowing ladder operators of order $l_{x}$ and $l_{y}$ that satisfy polynomial Heisenberg algebras (PHA's) ($i=x,y$)
\begin{equation}
  [H_{i},a_{i}^{\dagger}]=\lambda_{i}a_{i}^{\dagger},\quad [H_{i},a_{i}]=-\lambda_{i}a_{i}, \label{PHA1}
\end{equation}
\begin{equation}
  a_{i}a_{i}^{\dagger}=Q_{i}(H_{i}+\lambda_{i}),\quad a_{i}^{\dagger}a_{i}=Q_{i}(H_{i}), \label{PHA2}
\end{equation}
where $\lambda_{x}$ and $\lambda_{y}$ are constants. $Q_{x}$ and $Q_{y}$ are polynomials in $H_{x}$ and $H_{y}$. The second order integrals $K$ consists only in the difference of the 1D parts and the higher order intgrals $I_{-}$ and $I_{+}$ (with $n_{1}\lambda_{x}=n_{2}\lambda_{y}=\lambda$, $n_{1}$,$n_{2}$ $\in \mathbb{Z}^{*}$ ) take a very convenient factorized form
\begin{equation}
K=\frac{1}{2\lambda}(H_{x}-H_{y}),\quad I_{-}=a_{x}^{n_{1}}a_{y}^{\dagger n_{2}},\quad I_{+}=a_{x}^{\dagger n_{1}}a_{y}^{n_{2}}. \label{Int}
\end{equation}
The integrals are respectively of order 2, $l_{x}n_{1}+l_{y}n_{2}$ and $l_{x}n_{1}+l_{y}n_{2}$ and generate a polynomial algebra with three generators $PA(3)$.
\begin{equation}
  [K,I_{\pm}]=\pm I_{\pm}, \quad [I_{-},I_{+}]=F_{n_{1},n_{2}}(K+1,H)-F_{n_{1},n_{2}}(K,H),  \label{PolA}
\end{equation}
\begin{equation}
  F(K)=\prod_{i=1}^{n_{1}}Q_{x}\left(\frac{H}{2}+\lambda K-(n_{1}-i)\lambda_{x}\right)
  \prod_{j=1}^{n_{2}}Q_{y}\left(\frac{H}{2}-\lambda K+j\lambda_{y}\right), \label{Fn1n2}
\end{equation}

where $F(K)$ is a polynomial of degree $n_{1}l_{x}+n_{2}l_{y}$ in the generator $K$ and the polynomial algebra is of order $n_{1}l_{x}+n_{2}l_{y}-1$. As it is expected for a 2D systems it involved the central element $H$.

\subsection{Polynomial algebra of the four families}

Explicit form of the ladder operators ($c_{o}$,$c_{o}^{\dagger}$), ($c_{ro}$,$c_{ro}^{\dagger}$), ($c_{ko}$,$c_{ko}^{\dagger}$) and ($c_{kro}$,$c_{kro}^{\dagger}$) that are respectively ladder operators of the harmonic oscillator $H^{(o)}$, radial oscillator $H^{(ro)}$, k-step extension of the harmonic oscillator $H^{(ko)}$ and k-step extension of the radial oscillator $H^{(kro)}$ have been given and their corresponding PHA \cite{marquette14a}. The operators  $c_{ko}$ and $c_{kro}$ were build using deleting and creating approaches of SUSYQM.

Let us summarize by Table 1, the parameters of the PHA's and their order for each of the four 1D components involved in the four families of superintegrable systems. 

\begin{table}[h!]
\caption{ The values $|a_{x}|$, $|a_{y}|$, $|Q_{x}|$ and  $|Q_{y}|$ denote the order of the ladder operators and the order of the polynomials $Q_{x}$ and $Q_{y}$. The parameters $\lambda_{x}$ and $\lambda_{y}$ of the polynomial Heisenberg algebras (PHA's) are also given. \\}

\begin{center}
\begin{tabular}{|l|l|l|l|l|l|l|l|l|l|l|}
  \hline
 Families  & $\lambda_{x}$ & $\lambda_{y}$ & $a_{x}$ & $a_{y}$ & $|a_{x}|$ & $|a_{y}|$   & $Q_{x}$ & $Q_{y}$ & $|Q_{x}|$ & $|Q_{y}|$    \\[0.05cm]
  \hline
$H_{a}$ & $2m_{k}+2$ & 2   &   $c_{ko}$   & $c_{o}$ & $m_{k}+1$ & 1   & $Q_{ko}$    &  $Q_{o}$  & $m_{k}+1$   &   1      \\[0.05cm]
  \hline
$H_{b}$  & $2m_{k}+2$ & 2 & $c_{kro}$ &  $c_{o}$   & $2m_{k}+2$ & 1    & $Q_{kro}$   &  $Q_{o}$  & $2m_{k}+1$   &  1        \\[0.05cm]
  \hline
$H_{c}$  & $2m_{k}+2$ & 2  & $c_{ko}$   & $c_{ro}$  & $m_{k}+1$ & 2   & $Q_{ko}$    &  $Q_{ro}$  & $m_{k}+1$   &   2         \\[0.05cm]
  \hline
$H_{d}$  & $2m_{k}+2$ & 2  & $c_{kro}$  & $c_{ro}$  & $2m_{k}+2$ & 2   & $Q_{kro}$   &  $Q_{ro}$  & $2m_{k}+1$   &  2       \\[0.05cm]
  \hline 
\end{tabular}
\end{center}
\end{table}

These ladder operators allow to generate integrals of motion factorized forms as given by Eq.~(\ref{Int}) and the polynomial algebra from Eq.~(\ref{PolA}) and~(\ref{Fn1n2}) for the four families of superintegrable systems. We present in Table 2, the order of these integrals $I_{-}$ and $I_{+}$, the order of the polynomial $|F_{n_{1},n_{2}}|$ and the order of the polynomial aLgebra $|PA(3)|$. The integrals $K$ is of order 2 in all cases. Let us also present explicitly the polynomial algebra by provinding $F_{n_{1},n_{2}}(K)$ for the Hamiltonians $H_{a}$, $H_{b}$, $H_{c}$ and $H_{d}$ :
  
\begin{table}[h!]
\caption{The parameters $n_{1}$, $n_{2}$ and $\lambda$ for integrals of the four families of systems $H_{a}$, $H_{b}$, $H_{c}$ and $H_{d}$ are given. $|I_{+}|$, $|I_{-}|$, $|F_{n_{1},n_{2}}|$ and $|PA(3)|$ denote respectively the order of the integrals of motion $I_{+}$ and $I_{-}$, the order of the polynomial $F_{n_{1},n_{2}}$ and the order of the related polynomial algebra.\\}

\begin{center}
\begin{tabular}{|l|l|l|l|l|l|l|l|}
  \hline 
 Families & $n_{1}$ & $n_{2}$ & $\lambda$  & $|I_{+}|$ & $|I_{-}|$ & $|F_{n_{1},n_{2}}|$ & $|PA(3)|$  \\[0.05cm]
  \hline
$H_{a}$   & 1 & $m_{k}+1$  & $2m_{k}+2$   &  $2m_{k}+2$  &  $2m_{k}+2$  &  $2m_{k}+2$  &  $2m_{k}+1$   \\[0.05cm]
  \hline
$H_{b}$  & 1 & $m_{k}+1$ & $2m_{k}+2$  &  $3m_{k}+3$  & $3m_{k}+3$  &  $3m_{k}+2$ &  $3m_{k}+1$  \\[0.05cm]
  \hline
$H_{c}$  & 1 & $m_{k}+1$ & $2m_{k}+2$  &   $3m_{k}+3$  & $3m_{k}+3$  & $3m_{k}+3$ & $3m_{k}+2$ \\[0.05cm]
  \hline
$H_{d}$ & 1 & $m_{k}+1$ &  $2m_{k}+2$   &   $4m_{k}+4$  &  $4m_{k}+4$  & $4m_{k}+3$  & $4m_{k}+2$ \\[0.05cm]
  \hline 
\end{tabular}
\end{center}
\end{table}

\begin{equation}
F_{a}(K)=\prod_{n=1}^{k}(\frac{H}{2}+(2m_{k}+2)K + 2 m_{n}+1) \label{Fa}
\end{equation}
\[ \prod_{r=1, r \neq T}^{m_{k}}(\frac{H}{2}+(2m_{k}+2)K-2r -1) \prod_{j=1}^{m_{k}+1}(\frac{H}{2}-(2m_{k}+2)K+2j),\]

\begin{equation}
F_{b}(K)= \prod _{n=1}^{k}(\frac{H}{2}+(2m_{k}+2)K-\alpha + 2m_{n}-k+1) \label{Fb}
\end{equation}
\[ \prod_{r=0}^{m_{k}}(\frac{H}{2}+(2m_{k}+2)K +\alpha -2 r + k -1) \]
\[ \prod_{s=1, r \neq T}^{m_{k}}(\frac{H}{2}+(2m_{k}+2)K-\alpha-2s-k -1) \prod_{j=1}^{m_{k}+1}(\frac{H}{2}-(2m_{k}+2)K+2j),\]

\begin{equation}
F_{c}(K)=\prod_{n=1}^{k}(\frac{H}{2}+(2m_{k}+2)K + 2 m_{n}+1)\prod_{r=1, r \neq T}^{m_{k}}(\frac{H}{2}+(2m_{k}+2)K-2r -1) \label{Fc}
\end{equation}
\[ \prod_{j=1}^{m_{k}+1}\frac{1}{4}(\frac{H}{2}-(2m_{k}+2)K+2j-\frac{3}{2}-l)(\frac{H}{2}-(2m_{k}+2)K+2j-\frac{1}{2}+l),\]

\begin{equation}
F_{d}(K)= \prod _{n=1}^{k}(\frac{H}{2}+(2m_{k}+2)K-\alpha + 2m_{n}-k+1) \label{Fd}
\end{equation}
\[ \prod_{r=0}^{m_{k}}(\frac{H}{2}+(2m_{k}+2)K +\alpha -2 r + k -1)\prod_{s=1, r \neq T}^{m_{k}}(\frac{H}{2}+(2m_{k}+2)K-\alpha-2s-k -1) \]
\[ \prod_{j=1}^{m_{k}+1}\frac{1}{4}(\frac{H}{2}-(2m_{k}+2)K+2j-\frac{3}{2}-l)(\frac{H}{2}-(2m_{k}+2)K+2j-\frac{1}{2}+l),\]

where $T =\{ m_{k}-m_{k-1}, \ldots, m_{k}-m_{1} \}.$

\section{Direct approach of constructing unirreps}

The polynomial algebras given by Eq.~(\ref{PolA}) and~(\ref{Fn1n2}) with Eqs.~(\ref{Fa}),~(\ref{Fb}),~(\ref{Fc}) and~(\ref{Fd}) are deformed $u(2)$ Lie algebras and the finite-dimensional unirreps may be characterized by their basis states by $|N, \tau,s, \sigma \rangle$ where $\sigma=-s,-s+1,\ldots, s$ and $\tau$ distinguishes between repeated representations specified by the same $s$ ( integer or half-integer). Moreover, $\sigma$ denotes the eigenvalue of $I_{0}=K+c$, where $c$ is some representation-dependent constant and $I_{+}$, $I_{-}$ are such that $I_{+}|N, \tau,s,s \rangle=I_{-}|N, \tau,s,-s \rangle =0$. From these constraints, we obtain the value of $s$ for every $N$ value. In order to implement the method and obtain the unirreps, we need to denote the states differently. These states can be represented in unified way by the two quantum numbers $\nu_{x}$ and $\nu_{y}$ as $|\nu_{x} \rangle_{1} | \nu_{y} \rangle_{2}$ but also using the quantum numbers $\nu_{x}$ and $N$ , where $N=\nu_{x}+\nu_{y}+1$ as $|N, \nu_{x}  \rangle = |\nu_{x}  \rangle_{1} |N-\nu_{x}-1  \rangle_{2}$. The action of the integrals $I_{-}$ and $I_{+}$ in this new basis can be obtained from the one of the lowering and raising operators as $I_{-}$ and $I_{+}$ are in fact polynomials of the ladders operators. The action of the operators $c_{ko}$ or $c_{kro}$ in the x axis was shown to be \cite{marquette14a} :
\begin{equation}
\begin{split}
  & c\psi^{(2)}_{\nu} = 0, \qquad \nu=-m_k-1, \ldots, -m_1-1, 1, 2, \ldots, m_k-m_{k-1}-1,  \\
  & \qquad  m_k-m_{k-1}+1, \ldots, m_k-m_1-1, m_k-m_1+1, \ldots, m_k,  
\end{split}
\end{equation}
\begin{equation}
 c\psi^{(2)}_0 \propto \psi^{(2)}_{-m_k-1},\quad c\psi^{(2)}_{m_k-m_i} \propto \psi^{(2)}_{-m_i-1}, \qquad i=1, 2, \ldots, k-1, 
\end{equation}
\begin{equation}
 c\psi^{(2)}_{\nu} \propto \psi^{(2)}_{\nu-m_k-1}, \qquad \nu=m_k+1, m_k+2, \ldots.
\end{equation}  
and for $c_{o}$ and $c_{ro}$ in the y axis is given by
\begin{equation}
 c\psi^{(2)}_{0} = 0, \quad c\psi^{(2)}_{\nu} \propto \psi^{(2)}_{\nu+1}, \qquad \nu=1,2,3, \ldots \quad .
\end{equation} 
The key element here consists in the fact the action of the ladder operators of the harmonic oscillator and radial oscillator on their respective eigenstates is similar even if their order and their PHA are different. The same observation can be made for their k-step extension as the action of the k-step extension of the harmonic oscillator and the k-step extension of the radial oscillator on their respective eigenfunctions in terms of EOP are similar even if their order as differential operators differ greatly and their corresponding PHA. Thus the action of integrals in the $|N, \nu_{x}  \rangle$ basis is obtained in unified way for the four superintegrable systems with $K |N, \nu_{x} \rangle \propto |N, \nu_{x} \rangle$ and
\begin{equation}
  I_{+}  |N,\nu_{x} \rangle \propto |N,\nu_{x}+ m_{k}+1 \rangle, \quad I_{-}  |N,\nu_{x} \rangle \propto |N,\nu_{x}- m_{k}-1 \rangle.
\end{equation}
The zero modes of these operators are enumerated in term of the values of $\nu_{x}$ and $N$ by considering
\begin{equation}
I_{-} |N, \nu_{x}  \rangle = 0,\quad I_{+} |N, \nu_{x}  \rangle = 0.
\end{equation}
Tables of these states were presented in the paper \cite{marquette14a}. Starting from states annihilated by  $I_{-}$ and acting with $I_{+}$ until we reach a state annihilated by $I_{+}$ we identify the values of $s$ where $2s=p$ is the number of states in the sequences. Furthermore, the $\sigma$ is associated with each state forming a given sequence. Using notation $N= \lambda n_{1}n_{2}+\mu$ with appropriate values of $\lambda$ and $\mu$, we enumerate all the states and obtain the values of $\tau$. This construction is based on the action on physical states and thus there is no problem of nonphysical states as for the Daskaloyannis approach. All the states are classified in sequences of different length that correspond to finite dimensional unirreps and given in Table 3. There are 11 cases and 21 types of sequences. 

\begin{table}[h!]

\caption{Set of $p=2s$ values with their number of occurrences, number $\cal N$ of unirreps per level, and total level degeneracy for the polynomial algebra $PA(3)$ given by Eq.~(\ref{PolA}) and~(\ref{Fn1n2}) with Eqs.~(\ref{Fa}),~(\ref{Fb}),~(\ref{Fc}) and~(\ref{Fd}) related to Hamiltonians $H_a$, $H_b$, $H_c$, or $H_d$. In this table, $j$ runs over 2, 3, \ldots, $k-1$.\\}

\begin{center}
\begin{tabular}{|l|l|l|l|l|l|}
  \hline
Case &  $\lambda$ & $\mu$ & $2s=p$ & $\cal N$ & $\deg(E_N)$\\[0.1cm]
  \hline
 1 & $-1$ & $1, \ldots, m_{k}-m_{k-1}$ & 0 & 1 & 1 \\[0.1cm]
  \hline
 2 & $-1$ & $m_{k}-m_{j}+1,\ldots,m_{k}-m_{j-1}$ & $0^{k-j+1}$ & $k-j+1$ & $k-j+1$ \\[0.1cm]
  \hline
 3 & $-1$ & $m_{k}-m_{1}+1,\ldots, m_{k}$ & $0^{k}$ & $k$ & $k$  \\[0.1cm]
  \hline
 4 & 0 & 0 & $0^{k}$ & $k$ & $k$  \\[0.1cm]
  \hline
 5 & $0$ & $1,\ldots,m_{k}-m_{k-1}$ & $1$ & $\mu+k-1$ & $N+k$ \\[0.1cm]
   & & & $0^{\mu+k-2}$ & & \\[0.1cm]
  \hline
 6 & $0$& $m_{k}-m_{j}+1, \ldots,m_{k}-m_{j-1}$ & $1^{k-j+1}$ & $\mu+j-1$ & $N+k$ \\[0.1cm]
   &  &  & $0^{\mu-k+2j-2}$ & & \\[0.1cm]
     \hline
 7 & $0$& $m_{k}-m_{1}+1, \ldots,m_{k}$ & $1^{k}$ & $\mu$ & $N+k$ \\[0.1cm]
   &  &  & $0^{\mu-k}$ & & \\[0.1cm] 
     \hline
 8 & $1,2,\ldots$ & 0 & $\lambda^k$ & $m_k+1$ & $N+k$ \\[0.1cm]
   &  & & $(\lambda-1)^{m_k-k+1}$ & & \\[0.1cm]
     \hline
 9 & $1,2,\ldots$& $1,\ldots,m_{k}-m_{k-1}$ & $\lambda+1$ & $m_{k}+1$ & $N+k$ \\[0.1cm]
   &  &  & $\lambda^{\mu+k-2}$ & & \\[0.1cm]
   &  &  & $(\lambda-1)^{m_{k}-\mu-k+2}$ & & \\[0.1cm]
     \hline
 10 & $1,2,\ldots$& $m_{k}-m_{j}+1,\ldots,m_{k}-m_{j-1}$ & $(\lambda+1)^{k-j+1}$ & $m_{k}+1$ & $N+k$ \\[0.1cm]
    &  &  & $\lambda^{\mu-k+2j-2}$ & & \\[0.1cm]
    &  &  & $(\lambda-1)^{m_{k}-\mu-j+2}$ & & \\[0.1cm]
     \hline
  11 & $1,2,\ldots$& $m_{k}-m_{1}+1,\ldots,m_{k}$ & $(\lambda+1)^k$ & $m_{k}+1$ & $N+k$ \\[0.1cm]
     &  &  & $\lambda^{\mu-k}$ & & \\[0.1cm]
     &  &  & $(\lambda-1)^{m_{k}-\mu+1}$ & & \\[0.1cm]    
  \hline 
\end{tabular}
\end{center}

\end{table}

\section{Daskaloyannis approach of constructing unirreps}

Alternatively, the polynomial algebras given by Eq.~(\ref{PolA}) and~(\ref{Fn1n2}) with Eqs.~(\ref{Fa}),~(\ref{Fb}),~(\ref{Fc}) and~(\ref{Fd}) belong to the class of polynomial algebras of arbitrary order studied \cite{Isaac14} and for which constraints from Jacobi identity, the Casimir operators and the realizations as deformed oscillator algebras were obtained. A formula can be obtained for the structure function at any given order. In general the construction of unirreps involve the realization of the Casimir operators in terms of central elements i.e. $H$ for 2D Hamiltonians. In our case, we have a very specific structure and the integrals are build from ladder operators generated by various supercharges and these ladder operators generate PHA's. Due to these underlying structures the realization of the polynomial algebra with three generators as deformed oscillator algebra can be obtained in a straightforward manner by taking $b^{\dagger}=I_{+}$, $b=I_{-}$ and $N=K-u$ (where $u$ is a representation dependent parameter that is determined using further constraints) :
\begin{equation}
[N,b^{\dagger}]=b^{\dagger},\quad [N,b]=-b, 
\end{equation}
\begin{equation}
b^{\dagger}b=\Phi(N) ,\quad bb^{\dagger}=\Phi(N+1),\quad \Phi(H,u,N)=F_{n_{1},n_{2}}(K,H) .
\end{equation}
In general the Fock basis is $|c,n>$ with $C|c,n>=c|c,n>$ and $N|c,n>=n|c,n>$ ( where $C$ is the Casimir operator) and using realization of the Casimir operators in term of $H$ we can in fact use a basis $|E,n>$ such that $H|E,n>=E|E,n>$ and $N|E,n>=n|E,n>$. For our 4 systems, we can simply take a such basis $|E,n>$. Furthermore, we have $b|E,0>=0.$ and the action of operators $b$ and $b^{\dagger}$ are given by $b^{\dagger}|n>\propto|E,n+1>$ and $b|n>\propto|E,n-1>$. We have the existence of finite dimensional unitary representations if we impose the following three constraints : $\Phi(E,p+1,u)=0$, $\Phi(E,0,u)=0$ and $\Phi(E,n,u)>0$, $\forall \; n>0$. The energy $E$ and the constant $u$ can be obtained from this set of constraints that are algebraic
equations as 
\begin{equation*}
E_{i}=E_{i}(p), \quad \Phi_{i}(x)= x(p+1-x)\prod_{j}^{L}( \alpha_{i}^{(j)}p +\beta_{i}^{(j)} x +\gamma_{i}^{(j)}), \quad u_{i}=u_{i}(p).
\end{equation*}
For most cases, the solutions will be valid for $p \in \mathbb{N}$ and $p$ correspond to a degeneracy of $p+1$. However, truncated solutions that are valid only for $ p \leq p_{max}$ were observed \cite{marquette09a,marquette13a} and such phenomenon is associated with isolated multiplets of the ladder operators that are used to generate the integrals. However, even if the solutions are not truncated, $p+1$ is not necessarily the total number of degeneracies as for a given energy level $E(p)$ more than one solutions $E_{i}(p)$ can produce the same energy $E$ with a different values of p. The total number of degeneracies will be given by unions of finite dimensional unirreps.  Let us point out the fact that the structure function is a polynomial allows the existence of many solutions and the possibility to obtain algebraically these more complicated patterns. Let us present a calculation for the Case $H_{a}$ and with one 1-step. There are two families of unirreps
 with $p \in \mathbb{N}$, $k \in \{1, 2, \ldots, m+1\}$, $l \in \{1, 2, \ldots, m\}$, 
\begin{equation}
\Phi_1=  x \prod_{i=1}^m [(m+1)x - m - 1 - i] \prod_{j=1}^{m+1} [(m+1)(p+1-x) - m + j - k],
\end{equation}
\begin{equation}
 \Phi_2 = [(m+1) x + m + 1 + l] \prod_{i=1}^m [(m+1)x + l - i] \prod_{j=1}^{m+1} [(m+1)(p+1-x) + j - k].
\end{equation}
The associated energy spectrum is given by  
\begin{equation}
  E_1(p) = 2 [(m+1)p +1 - k],\quad 2(m+1) u_1 = - \frac{E}{2} - 2m - 1,
\end{equation}
\begin{equation}
  E_2(p) = 2 [(m+1)(p+1) + l - k + 1], \quad 2(m+1) u_2 = - \frac{E}{2} + 2l + 1,
\end{equation}
where $p \in \mathbb{N}$. We set $\nu_x = (m+1) n_x + a_1$, $\nu_y = (m+1) n_y + a_2$, with $n_x, n_y \in \mathbb{N}$, $a_1 \in \{-m-1, 1, 2, \ldots, m\}$, and $a_2 \in \{0, 1, \ldots, m\}$. The energy $E$ obtained from separation of variables and wavefunctions involving Hermite EOP of type III can be rewritten as $E = 2[(m+1)(n_x+n_y) + a_1 + a_2 + 1]$. $E_1$ and $E_2$, correspond to $E$ with $n_x + n_y = p \in \mathbb{N}$, $a_2 = m+1-k \in \{0, 1, \ldots, m\}$, and $a_1 = -m-1$ or $a_1 = l \in \{1, 2, \ldots, m\}$. The total number of degeneracies is not obtained directly as these $(m+1)^2$ solutions can provide the same energy for different value of $p$ and one needs to enumerate the states from these expressions ( i.e. $E_{1}(p)$ and $E_{2}(p)$ ) and taking into account that an energy with a given $p$ represent $p+1$ states.There are only 5 different cases and 9 types of sequences \cite{marquette13b} that are contained in the more general results for k-step extensions and presented in Table 3.

\section{Conclusion}

We presented four new families of systems involving k-step extension of the harmonic oscillator and radial oscillator related to Hermite and Laguerre EOP of type III. Let us denote, that families that are superintegrable and involve k-step extension in both axis were also obtained and studied \cite{marquette14a}. These four superintegrable models are caracterised by interesting degeneracies patterns that are related with non trivial finite dimensional unirreps given by Table 3. These results demonstrate how the symmetry algebra and its representations explain the degeneracies of these systems as observed in other superintegrable Hamiltonians with less complicated pattern. These results also showed that superintegrable systems with spectrum  beyond isotropic and anisotropic harmonic oscillator type of degeneracies exist and can be build by using these 1D components related to type III EOP.

Furthermore, we showed that the Daskaloyannis approach can be used for these systems with higher order integrals, however appropriate set of integrals generated from ladder operators without multiplets isolated from excited states need to be used. We pointed out that a direct approach can be taken to study in an unified way the four quantum models by using the action of the integrals. Moreover, this novel approach relying on action of the integrals also reveals the underlying structure of the unirreps in a more explicit way in terms of states than the Daskaloyannis approach that is very convenient when no factorized form in terms of ladder have been identified for the integrals. An open problem is the generalisation of these approaches for higher dimensional superintegrable quantum systems and their related finitely generated higher rank polynomial algebras which possess more complicated commutation relations \cite{daska11}.

\ack The research of I.M. was supported by a Discovery Early Career Researcher Award DE130101067.

\smallskip

\end{document}